\documentclass[pra,amsmath,amssymb,twocolumn,footinbib,superscriptaddress]{revtex4}

\usepackage{amsmath,amssymb}
\usepackage[usenames]{color}
\usepackage{amssymb}
\usepackage{dsfont}
\usepackage{grffile}
\usepackage[pdftex]{graphicx}
\usepackage{amsmath, amstext, amssymb, amsfonts, amsxtra}
\usepackage{textcomp}
\usepackage{xspace}
\usepackage{hyperref}
\DeclareSymbolFontAlphabet{\amsmathbb}{AMSb}

\newcommand{\cur}{\mathcal{J}}

\newcommand{\Rec}{\mathcal{R}}

\newcommand{\rhop}{\hat{\rho}}

\newcommand{\Hop}{\hat{H}}

\newcommand{\Dop}{\mathcal{D}}

\newcommand{\im}{{\rm i}}

\newcommand{\tun}{t}

\definecolor{john}{rgb}{1,0.1,0.1}

\newcommand{\sutd}{Science and Math Cluster and EPD Pillar, Singapore University of Technology and Design, 8 Somapah Road, 487372 Singapore.}
\newcommand{\bath}{Department of Physics, University of Bath, Claverton Down, Bath BA2 7AY, U.K.}
\newcommand{\bristol}{H.H. Wills Physics Laboratory, University of Bristol, Bristol BS8 1TL, UK.}
\newcommand{\hamburg}{Max Planck Institute for the Structure and Dynamics of Matter,
University of Hamburg CFEL, Hamburg, Germany.}
\newcommand{\ireland}{School of Physics, Trinity College Dublin, Dublin 2, Ireland.}

\begin{document}

\title{Heat current rectification and mobility edges}

\author{Vinitha Balachandran}
\affiliation{\sutd} 
\author{Stephen R. Clark} 
\affiliation{\bristol}
\affiliation{\bath}
\affiliation{\hamburg} 
\author{John Goold}
\affiliation{\ireland} 
\author{Dario Poletti}
\affiliation{\sutd}

\begin{abstract}
We investigate how the presence of a single-particle mobility edge in a system can generate strong heat current rectification. Specifically, we study a quadratic bosonic chain subject to a quasi-periodic potential and coupled at its boundaries to spin baths of differing temperature. We find that rectification increases by orders of magnitude depending on the spatial position in the chain of localized eigenstates above the mobility edge. The largest enhancements occur when the coupling of one bath to the system is dominated by a localized eigenstate, while the other bath couples to numerous delocalized eigenstates. By tuning the parameters of the quasi-periodic potential it is thus possible to vary the amplitude, and even invert the direction, of the rectification.
\end{abstract}
%


\maketitle

{\it Introduction:} The possibility to control heat transport at the nano-scale can open a large number of opportunities \cite{BenentiWhitney2017}. This has motivated a large number of studies, both at the classical and quantum level. One important class of systems studied is that of current rectifiers, which are systems in which the magnitude of the resulting current is very different depending on the direction of current induced by the external bias, e.g. analogous to well known diodes for electrical currents. 

In classical systems it was shown that coupled non-linear chains can be used to rectify heat flow thanks to a bias dependent mismatch of the spectral response of the chains \cite{TerraneoCasati2002, LiCasati2004} (see Ref.~ \cite{LiLi2012} for a review). Rectification has been observed also in quantum interacting chains \cite{Landi2014, Landi_many, LifaLi2009, ZhangLi2009, WerlangValente2014, BalachandranPoletti2018b}, and in particular, it was recently shown that a perfect spin current rectifier could be produced in segmented spin chains once the interaction exceeds a critical value \cite{BalachandranPoletti2018}. 

However, interactions are not necessary to obtain rectification. When magnetic fields are present in the baths, breaking time-reversal symmetry, quadratic spin chains can display heat rectification \cite{ArracheaAligia2009}. Particularly relevant for our work are the investigations done in Refs.~\cite{WuSegal2009, WuSegal2009b}. There they presented two sufficient conditions for the emergence of rectification: the first one is the presence of a mismatch in energy dependence of the density of states between the baths; the second condition, and also the most relevant for our work, is the presence of baths which, while of identical nature, have particles/excitations with different quantum statistics from that of the system which connects them.   

Recent years have also witnessed a significant interest in disorder and quasi-periodic systems (i.e. systems with an incommensurate potential). For one dimensional non-interacting quantum systems, any amount of disorder induces localization \cite{Anderson1958}, and localization can still be found  in interacting systems \cite{BaskoAltshuler2006, AbaninSerbyn2018}. Localization occurs also for quasi periodic potentials, as the prototypical Aubry-Andr\'e-Harper model, but only once the magnitude of the potential is larger than a certain threshold value \cite{AubryAndre1980, Harper1955}. 
The localizing effects of disordered and quasi-periodic potentials, in non-interacting and interacting cases, have been verified experimentally \cite{BillyAspect2008, RoatiInguscio2008, DErricoModugno2014, SchreiberBLoch2015, SmithMonroe2016, BordiaSchneider2016, BordiaBloch2017, RoushanMartinis2017, ChoiGross2016, BordiaBloch2017b, RubioAbadalGross2018, KohlertAidelsburger2018}.        

Systems with a quasi-periodic potential may present mobility edges, which means that there is a particular value of the energy which differentiates energy eigenstates which are localized from delocalized ones \cite{early_sarma,Biddle2010,DengPixley2017, LiDasSarma2016, LiDasSarma2015, ModakMukherjee2015, GaneshanDasSarma2015,calabria}, as recently observed experimentally \cite{LuschenBloch2018}. The transport properties of systems with mobility edges have been studied before, showing, for example, a transition between ballistic transport to insulating behavior separated by a critical line with subdiffusive transport \cite{PurkayashtaKulkarni2018}. 

In this work we consider two identical spin baths of differing temperature connected to the boundaries of a quadratic bosonic chain with a generalized Aubry-Andr\'e-Harper potential which induces a mobility edge. Since the baths and the system have different statistics, and the quasi-periodic potential breaks the spatial reflection symmetry, this model possesses the key ingredients for the occurrence of rectification \cite{WuSegal2009, WuSegal2009b}. We will show that the presence of a bulk mobility edge can in fact result in {\em strong} rectification when localized eigenstates cluster at one edge of the system connected to one of the baths. Consequently, drastically different non-equilibrium steady states (NESS) are generated in forward or reverse bias of the applied bath temperatures. The direction of the rectification can be controlled by tuning the quasi-periodic potential parameters that shift the spatial position of the localized modes. While the rectification is strongest at large temperature differences, we show the robustness of the effect by studying the rectification at different temperature differences. 
           
{\it Model:} 
We study a generalized Aubry-Andr\'e-Harper model consisting of a one dimensional lattice of $L$ non-interacting local bosonic modes with on-site modulation $V_{l}$ described by the Hamiltonian
\begin{align}\label{ham}
 H=-\tun\sum_{l=1}^{L-1}(a _{l}^{\dagger}a_{l+1}+a_{l+1}^{\dagger}a_{l})+\sum_{l=1}^{L}V_{l}a_{l}^{\dagger}a_{l},
 \end{align} 
where the operators $a_l$ ($a_l^\dagger$) annihilates (creates) a boson at site $l$. Furthermore, $\tun$ is the hopping parameter, $V_{l}=\mu+2\lambda \frac{1-\mathrm{cos}(2\pi l b+\phi)}{1+\alpha \mathrm{cos}(2\pi l b+\phi)}$ is the potential and it is characterized by the deformation parameter $\alpha$, on-site modulation strength $\lambda$, period $1/b$, phase parameter $\phi$ and a constant off-set value $\mu=2t$. To have
quasi-periodic modulation, $b$ is chosen to be irrational and we use $b=(\sqrt{5}-1)/2$. The case $\alpha=-1$ corresponds to a constant on-site energy $2\lambda$, whereas $\alpha=0$ is a rescaled version of the Aubry-Andr\'e-Harper model. Diagonalization of the Hamiltonian in Eq.~(\ref{ham}) gives $H = \sum_k \epsilon_k \eta^\dagger_k\eta_k$, where $\epsilon_k$ is the energy of the $k$th single-particle eigenstate and $\eta_k$ ($\eta^\dagger_k$) being its corresponding eigenmode annihilation (creation) operator. Local modes are expressed in terms of eigenmodes via a unitary  transformation $S$ as $a_l=\sum_kS_{l,k}\eta_k$. This model exhibits a single-particle mobility edge where eigenstates with an energy greater than $E_{\rm mob} = 2 \lambda\;(|t/\lambda|-1)/\alpha +\mu$ are localized for any value of $\phi$ \cite{GaneshanDasSarma2015}. In the following we will work in units for which the tunneling $\tun$, the Boltzmann constant $k_B$ and reduced Planck constant $\hbar$ are set to unity.     

The chain is coupled at its edges, denoted as site $\ell = \{1,L\}$, to identical heat baths consisting of non-interacting spins at different temperatures. For both baths $\ell$ the $\nu$th spin has an energy $\varepsilon_\nu$ and couples to the system with a strength $g_\nu$ via a term $\sum_{\nu}g_{\nu}(a_\ell+a_\ell^\dagger)(\sigma^+_{\nu,\ell}+\sigma^-_{\nu,\ell})$ where $\sigma^+_{\nu,\ell}$ and $\sigma^-_{\nu,\ell}$ are respectively the raising and lowering operators for the $\nu$th spin coupled to site $\ell$. The evolution of the system's density matrix $\rho$ in time $\tau$ in the presence of the heat baths is modeled by a Lindblad master equation \cite{Lindblad,GoriniSudarshan,BreuerPetruccione2007, supp}
\begin{align}\label{mastereq}
  \frac{d\rho}{d \tau}&=-\im[\Hop,\rho]\nonumber \\ 
  &+\sum_{k,\ell} |S_{\ell,k}|^2 J(\epsilon_k)\left[ n_S(\beta_\ell \epsilon_k)\left( \eta_k \rho \eta_k^\dagger -1/2 \{\eta_k^\dagger\eta_k,\rho\} \right) \right. \nonumber \\ 
   &+\left. \left(n_S(\beta_\ell \epsilon_k)+1\right) \left(\eta_k^\dagger \rho \eta_k -1/2 \{\eta_k\eta_k^\dagger,\rho\} \right) \right], 
\end{align}
where $\beta_\ell=1/T_\ell$ is the inverse temperature, $n_S(\beta_\ell\epsilon_k)=\left(e^{\beta_\ell\epsilon_k}+1\right)^{-1}$ is the spin occupation factor of the bath coupled to the boundary site $\ell$, and $J(\epsilon)=\sum_{\nu}\pi |g_{\nu}|^2\delta(\epsilon-\varepsilon_\nu)$ is the spectral density of the baths. We consider an ohmic spectral density for the baths so $J(\epsilon) \propto \epsilon$.          

For the NESS, the single particle density matrix is 
\begin{align}
     \langle \eta_{k}^{\dagger}\eta^{}_{k} \rangle &=   \frac{\sum_{\ell}|S_{\ell k}|^{2}n_S(\beta_l\epsilon_{k})}{ \sum_{\ell}|S_{\ell k}|^{2}\left[ 1-\zeta n_S(\beta_l\epsilon_{k})\right]}, \label{eta_dens}
\end{align}
with $\zeta = 2$ additionally reflecting the spin nature of the baths.
Note that if only one bath is coupled to the system then since $n_S(\beta\epsilon_k)/(1-2n_S(\beta\epsilon)) = n_B(\beta\epsilon) = (e^{\beta\epsilon_k}-1)^{-1}$ we recover the Bose occupation factor for each eigenstate $k$ with non-zero coupling to the bath, as expected.
The steady state heat current $\cur$ is then      
\begin{align}\label{curr}
\cur=&~\mathrm{Tr}\big[\Hop\Dop_1(\rhop)\big]=-\mathrm{Tr}\big[\Hop\Dop_L(\rhop)\big],\\ \nonumber
    =&\sum_{k}\epsilon_{k}|S_{1 k}|^{2} |S_{L k}|^{2}J(\epsilon_{k})\frac{n_S(\beta_1\epsilon_{k})-n_S(\beta_L\epsilon_{k})}{ \sum_{\ell}|S_{\ell k}|^{2}[1-\zeta n_S(\beta_\ell\epsilon_{k})]}. 
\end{align} 
We refer to {\it forward bias} ($\cur_f$) as the case in which the hotter bath, with temperature $T_h$, is coupled to the first site and the cold one, with temperature $T_c$, is coupled to the last, while {\it reverse bias} ($\cur_r$) is the opposite case. The magnitude of the rectification is signalled by the rectification coefficient, which is the ratio between the current in forward bias $\cur_{f}$ and that in reverse bias $\cur_{r}$
 \begin{equation}\label{rect2}
  \Rec=-\frac{\cur_{f}}{\cur_{r}}.
 \end{equation}
The rectification coefficient $\Rec =1$ when there is no rectification, while $\Rec \gg 1$, or $\Rec \ll 1$, signal strong rectification in one or the other direction. Importantly, if the baths had been bosonic the expression for the current in Eq.~(\ref{curr}) would be identical except for replacing $n_S$ with $n_B$ and setting $\zeta = 0$. We therefore observe that when the bath and system statistics are identical the current is antisymmetric in the exchange of $\beta_1$ with $\beta_L$, so there is no rectification, confirming the result in Refs.~\cite{WuSegal2009, WuSegal2009b}.       

\begin{figure}
\includegraphics[width=.9\columnwidth]{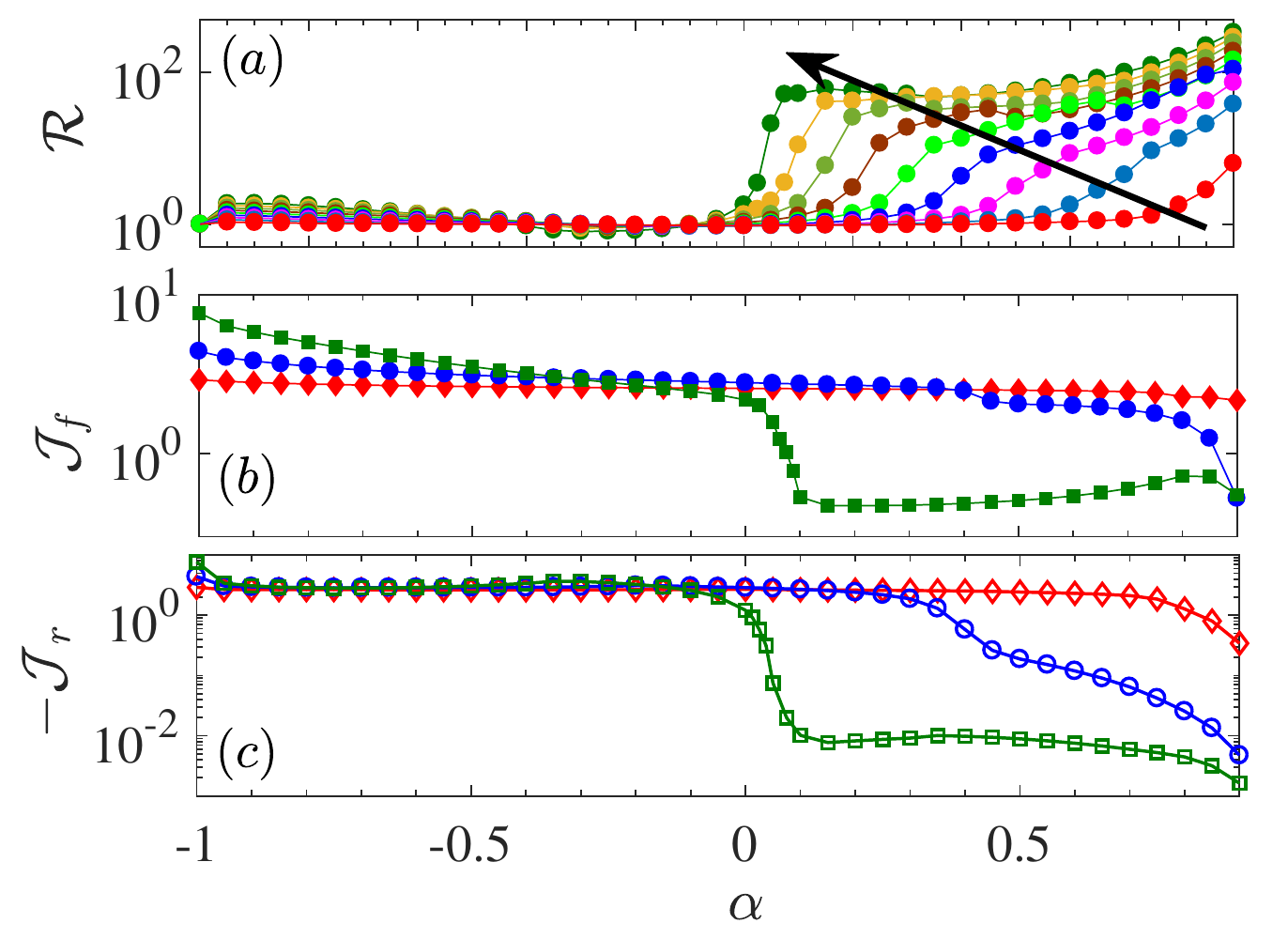}
\caption{(a) Rectification $\Rec$, (b) forward bias current $\cur_f$ and (c) reverse bias current $\cur_r$ versus the deformation parameter $\alpha$. In (a) different lines correspond to different values of $\lambda$, from $0.1$ to $0.9$ with steps of $0.1$, increasing in the direction of the arrow. (b)-(c) Forward and reverse bias currents for $\lambda=0.1$ (red $\diamond$), $0.4$ (blue $\circ$) and $0.9$ (green $\square$). $\cur_f$ is represented by full symbols while $\cur_r$ by empty symbols. Other parameters are: chain length $L=1000$, phase $\phi=\pi$ and temperatures of the baths are $T_h=1000.1$ and $T_c=0.1$. }
\label{fig:Fig1}
\end{figure}

{\it Results:} 
When $\lambda=0$ or $\alpha=-1$, the system corresponds to a uniform tight binding model, hence the forward and reverse currents are identical and there is no rectification. In Fig.~\ref{fig:Fig1}(a) we report the rectification coefficient $\mathcal{R}$ with $\alpha$ for a sequence of increasing $\lambda$'s. We observe that there are regions in the parameters space with very large rectifications, close to $\Rec\approx 100$. Moreover, as $\lambda$ increases, the range of $\alpha$'s for which this strong rectification appears increases significantly. In the limit of $\lambda=1$, there is strong rectification for all $\alpha>0$, but beyond that value all states are localized and the system is an insulator in both directions. For negative $\alpha$ instead, and as long as $\lambda$ is positive, all states are delocalized and the rectification is small. In Figs.~\ref{fig:Fig1}(b)-(c) we show the currents in the forward and reverse bias, respectively. The regime of high rectification corresponds to a reduction in both the forward and reverse bias currents, but with the suppression substantially larger for the reverse bias. 

\begin{figure}
\includegraphics[width=.85\columnwidth]{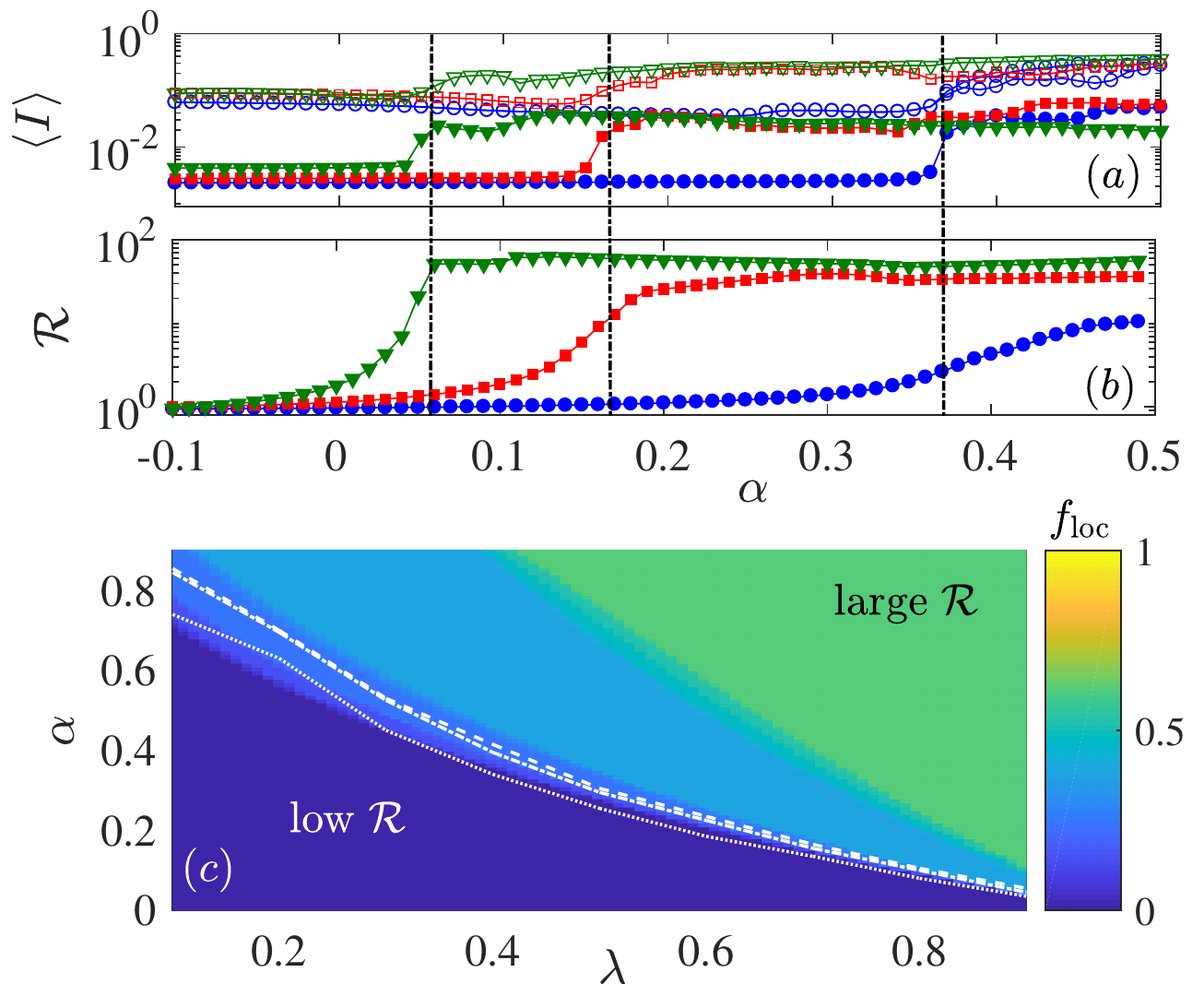}
\caption{(a) Inverse participation ratio $\langle I \rangle$ for forward (filled symbols) and reverse (empty symbols) bias versus deformation parameter $\alpha$ for $\lambda = 0.4$ (blue $\circ$), $0.7$ (red $\diamond$) and $0.9$ (green $\square$). (b) Rectification $\Rec$ versus $\alpha$ for $\lambda = 0.4$ (blue $\circ$), $0.7$ (red $\diamond$) and $0.9$ (green $\square$). In panels (a)-(b) we have used $T_h=1000+T_c$. 
(c) Density plot of fraction of localized states $f_{\rm loc}$ as a function of $\alpha$ and $\lambda$. The white lines represent the values of $\alpha$ and $\lambda$ at which the rectification varies significantly. More precisely we consider $T_h=10.1$ (dotted line), $T_h=100.1$ (dot-dashed line) and $T_h=1000.1$ (dashed line). Common parameters are $L=1000$, $\phi=\pi$ and $T_c=0.1$. Black dot-dashed lines highlight the value of $\alpha$ at which the inverse participation ratio increases significantly.}
\label{fig:Fig2}
\end{figure}

In order to understand the role of the mobility edge, we study the rectification together with localization properties of the NESS. This is most easily revealed by the inverse participation ratio $\langle I \rangle=\sum_k I(k)\langle \eta_k^\dagger\eta^{}_k\rangle$ where $\langle \eta_k^\dagger\eta^{}_k\rangle$ is given by Eq.~(\ref{eta_dens}) and $I(k)=\sum_{l}|S_{lk}|^{4}/\sum_{n}|S_{nk}|^{2}$. The inverse participation ratio $I(k)$ for an eigenstate $k$ is closer to unity the more localized the eigenstate is, while it is of the order of $1/L$ for a delocalized eigenstate. Correspondingly, a NESS with localized states significantly occupied will have a larger $\langle I\rangle$. In Fig.~\ref{fig:Fig2}(a) we show $\langle I\rangle$ for the forward (filled symbols) and reverse (empty symbols) biases, and for different $\lambda$'s. For these parameters, $\langle I\rangle$ for the reverse bias NESS is always larger than that of the forward bias. More importantly, at a particular value of $\alpha$, which changes with $\lambda$, both forward and reverse $\langle I\rangle$ grow significantly. As highlighted in Fig.~\ref{fig:Fig2}(b) by dot-dashed lines, it is around these values of $\alpha$ that we also observe a significant increase in the rectification.  

The relevance of the mobility edge in affecting the rectification can also be inferred in Fig.~\ref{fig:Fig2}(c). Here we show a density plot of the fraction of localized single-particle eigenstates $f_{\rm loc}$, where a state is considered localized if its energy is larger than the mobility edge $E_{\rm mob}$ \cite{GaneshanDasSarma2015}. For several different $T_h$ we plot on top of this the mid-point $\alpha$ where a significant uplift in $\Rec$ occurs as a function of $\lambda$, giving curves that delineates the low and high rectification regimes. These curves clearly demonstrate that the $\alpha$'s where rectification increases correspond closely with the values where there is a marked increase of localized states in the Hamiltonian. For $\lambda\approx 1$ we see large rectification for small $\alpha$ that is essentially independent of temperature, owing to the small energy threshold for localization. For smaller values of $\lambda$ a weak dependence on temperature is observed since $\Rec$ displays a broader slope, making the cross-over less sharp, and there is a decreased sensitivity to the mobility edge for smaller $T_h$ as the highest energy eigenstates are less populated.

\begin{figure}                  
\includegraphics[width=.9\columnwidth]{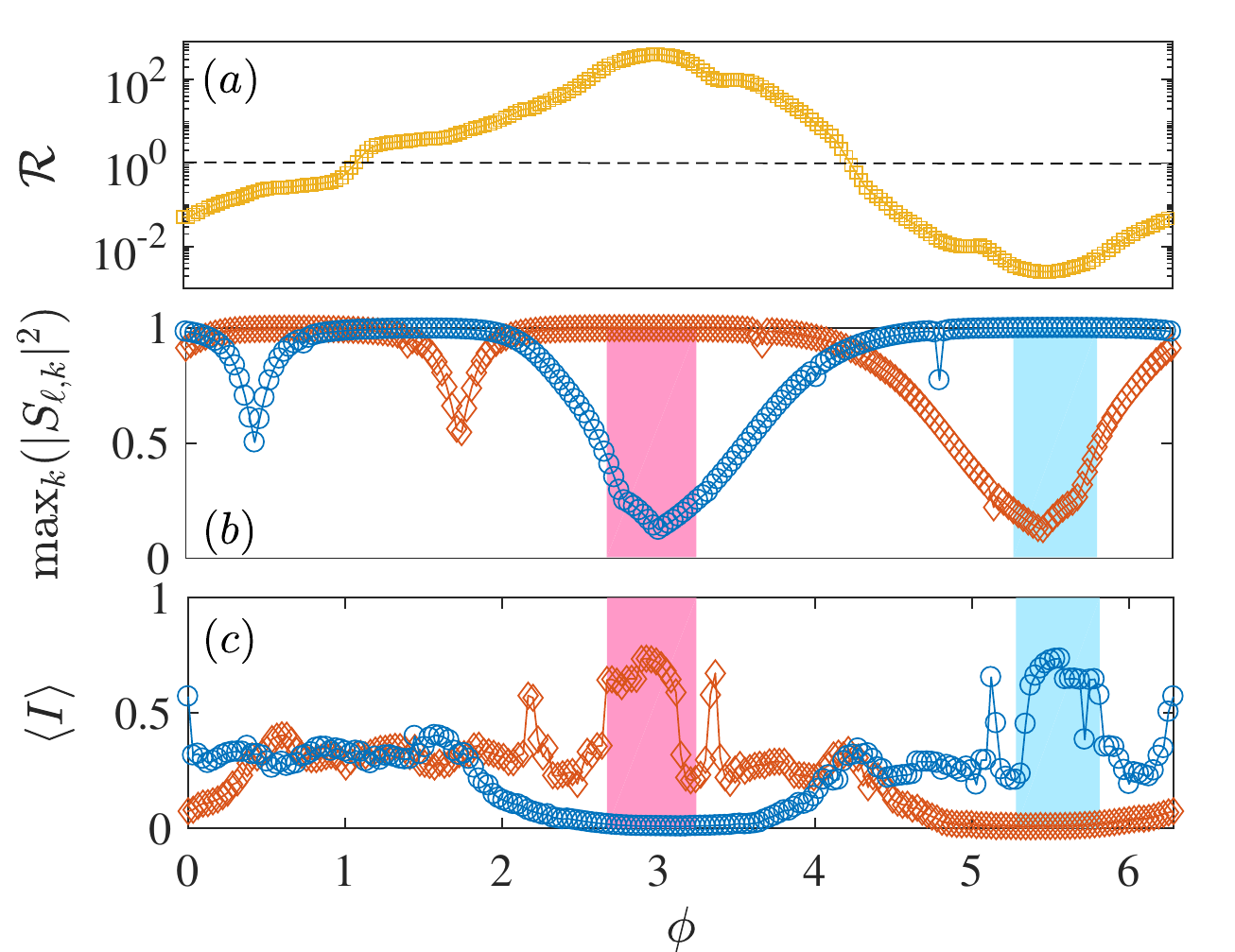}
\caption{ (a) Rectification versus phase $\phi$ in log-lin scale so highlight the regions in which the rectification is in different directions. (b) Maximum coupling magnitude of any $k-$mode to the first site (blue $\circ$) or to the last site (red $\diamond$). (c) Average inverse participation ratio $\langle I \rangle$ as a function of $\phi$ for the forward (blue $\circ$) and reverse (red $\diamond$) bias. Parameters are $\alpha=\lambda=0.9$, $T_h=1000.1$, $T_c=0.1$.}
\label{fig:Fig3}
\end{figure}

The link between rectification and localization is further unravelled by examining more closely the dependence of the rectification with the phase $\phi$ and the coupling of the baths to the system eigenstates. Fig.~\ref{fig:Fig3}(a), which depicts $\Rec$ as a function of $\phi$, shows that by tuning $\phi$ it is possible to obtain rectification around $\Rec \approx 400$ for $\phi\approx 3$, and $\Rec\approx 1/400$ (i.e. strong rectification in the opposite direction) for $\phi\approx 5.5$. Hence the potential parameter $\phi$ can be used to control the direction of the rectification. In Fig.~\ref{fig:Fig3}(b) we report $\max_k |S_{\ell,k}|$, i.e. the maximum coupling strength of the bath at $\ell$ to any system eigenstate $k$. Owing to the unitarity of $S$ we have that $0 \leq \max_k |S_{\ell,k}| \leq 1$. Consequently, when $\max_k |S_{\ell,k}|$ approaches unity it indicates the coupling is dominated by one eigenstate, which for the spatially localized system-bath interaction assumed here can only occur if that eigenstate is similarly localized. In contrast a small value indicates couplings to numerous delocalized eigenstates. We see that $\Rec \approx 1$ whenever the maximum coupling is similar at both edges. We note that rectification $\Rec > 1$ (or inversely $\Rec < 1$) arises whenever one bath couples to a single eigenstate, while the other couples to many. However, our key finding is that strong rectification $\Rec \gg 1$ (or $\Rec \ll 1$) manifests when one bath couples to a single eigenstate which is highly localized at the boundary, while the coupling for the other bath is spread over many delocalized eigenstates. This is demonstrated by the two disjoint shaded regions (pink for the bath at $\ell=L$ and light blue for $\ell=1$) in Fig.~\ref{fig:Fig3}(b)-(c), which signify the phases $\phi$ where $I(k) \approx 1$ for the maximally coupled eigenstate. In Fig.~\ref{fig:Fig3}(c) we show the average inverse participation ratio $\langle I \rangle$ of the NESS in forward (blue $\circ$) and reverse bias (red $\diamond$). Again we observe that in the regions of strongest rectification the NESS in one bias is highly delocalized, while it is strongly localized in the opposite bias.    

Strong rectification is therefore a consequence of this disparity in bath eigenstate couplings $S_{\ell,k}$, combined with the difference in statistics of the baths and system that is reflected in the form of the denominator of Eq.~(\ref{curr}). To understand this intuitively, suppose $T_c=0$ temperature while $T_h = \infty$ so that $n_S(\beta_\ell \epsilon_k)\approx 1/2$ for the modes coupled to the hot bath and $n_S(\beta_\ell\epsilon_k)=0$ for the cold bath. In this scenario Eq.~(\ref{curr}) gives a significantly larger current when the cold bath is coupled to a localized mode compared to when the hot bath is coupled to a localized mode \cite{simplemodel}. 

\begin{figure}
\includegraphics[width=.9\columnwidth]{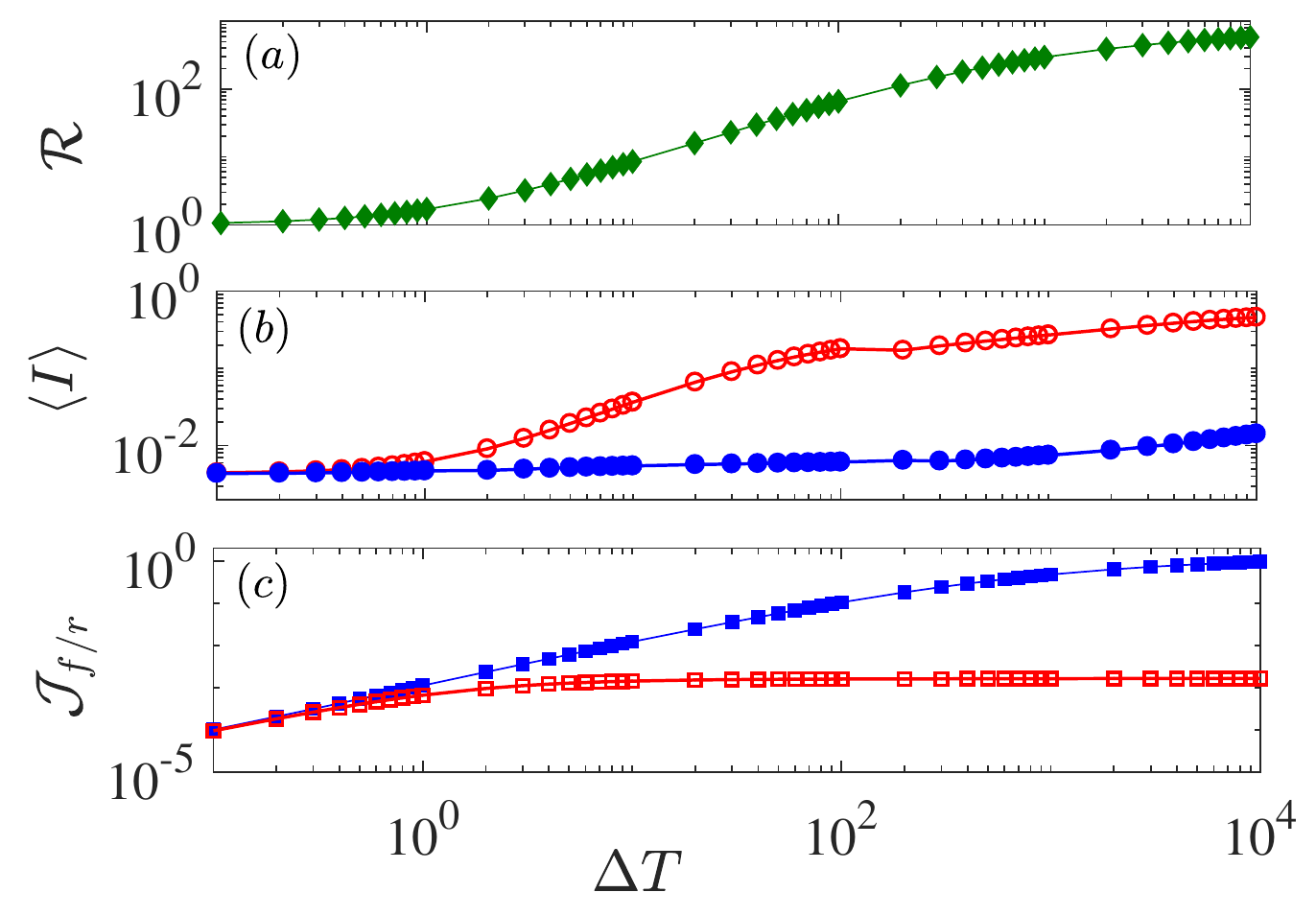}
\caption{(a) Rectification $\Rec$ versus temperature difference $\Delta T$. (b) Inverse participation ration $\langle I \rangle$ and (c) steady state currents for forward $\cur_f$ (filled blue symbols) and reverse $\cur_f$ (empty red symbols) biases, versus $\Delta T$. Common parameters are: $\alpha=\lambda=0.9$, $L=1000$, $\phi=\pi$ and $T_c=1$.}
\label{fig:Fig4}
\end{figure}

The rectification effect outlined here is observed over a wide range of temperatures. When temperature increases there are two main contributions: the bias which drives the current increases, so the current can increase, and the population of localized higher energy eignmodes also increases. Given the different occupation of localized and delocalized modes between the forward and the reverse bias, rectification increases with larger temperature difference $\Delta T$. We analyse this in Fig.~\ref{fig:Fig4}. Specifically, we show the rectification $\Rec$, average inverse participation ratio $\langle I\rangle$, and the forward and reverse currents $\cur_{f/r}$, as a function of $\Delta T$ in Figs.~\ref{fig:Fig4}(a)-(c), respectively.  For the reverse bias (empty symbols in panels (b)-(c)), $\langle I\rangle$ saturates to a larger value at lower $\Delta T$ compared to the forward bias (full symbols in panels (b)-(c)). Consequently, a large gap between $\cur_f$ and $\cur_r$ opens up as $\Delta T$ increases and the rectification grows to $\Rec \approx 580$. While large rectification occurs for such extreme temperature differences, Fig.~\ref{fig:Fig4}(a) nonetheless shows sizeable rectifications for much lower temperatures. Further analysis on how the phase $\phi$ and of the bath temperatures affect the localization and transport properties of the steady state can be found in \cite{supp}. 

{\it Conclusions:} 
We have studied the ability of a quadratic bosonic system with a mobility edge coupled to spin baths to rectify heat current. While the difference in particle statistics is fundamental in order to achieve rectification, we found regimes of strong rectification and identified its emergence as a result of one bath being strongly coupled to a highly localized mode while the other bath couples broadly to many delocalized modes. Strong rectification thus emerges due to the presence of a mobility edge, and it can be tuned by shifting the mobility or by tuning the location of the strongly localized modes. 

Crucially, since the system is ballistic, the current accompanying this large rectification coefficient remains appreciable even as the system size increases, in contrast for example to diffusive systems where the temperature gradients decrease with $L$. A possible proof of principle experimental implementation would be a chain of evanescently coupled cavities with the boundary cavities containing atomic ensembles mimicking the spin baths~\cite{Badshah2014,Norcia2018}. The effect observed should occur for other systems with different statistics in the bath and the system and mobility edges, such as a quasi-periodically modulated $XX$ spin-chain coupled to bosonic baths. Future work includes studying this and the coupling to the baths in more detail, as well as considering the role of many-body interactions.   

{\em Acknowledgments}: D.P. and V.B. are grateful to G. Benenti for insightful discussions. D.P. and V. B. acknowledge support Singapore Ministry of Education AcRF
MOE Tier-II (project MOE2016-T2-1-065, WBS R-144-000-350-112). SRC gratefully acknowledges support from the UK's Engineering and Physical Sciences Research Council (EPSRC) under grant No. EP/P025110/1. JG is supported by an SFI-Royal Society University Research Fellowship. JG acknowledges that this project has received funding from the European Research Council (ERC) under the European Union’s Horizon 2020 research and innovation programme (grant agreement No [758403]).              


\newpage 

\renewcommand{\thefigure}{A\arabic{figure}}
\setcounter{figure}{0}

\begin{appendix}

\section{Stability of the rectification to the phase parameter}

\begin{figure}
\includegraphics[width=\columnwidth]{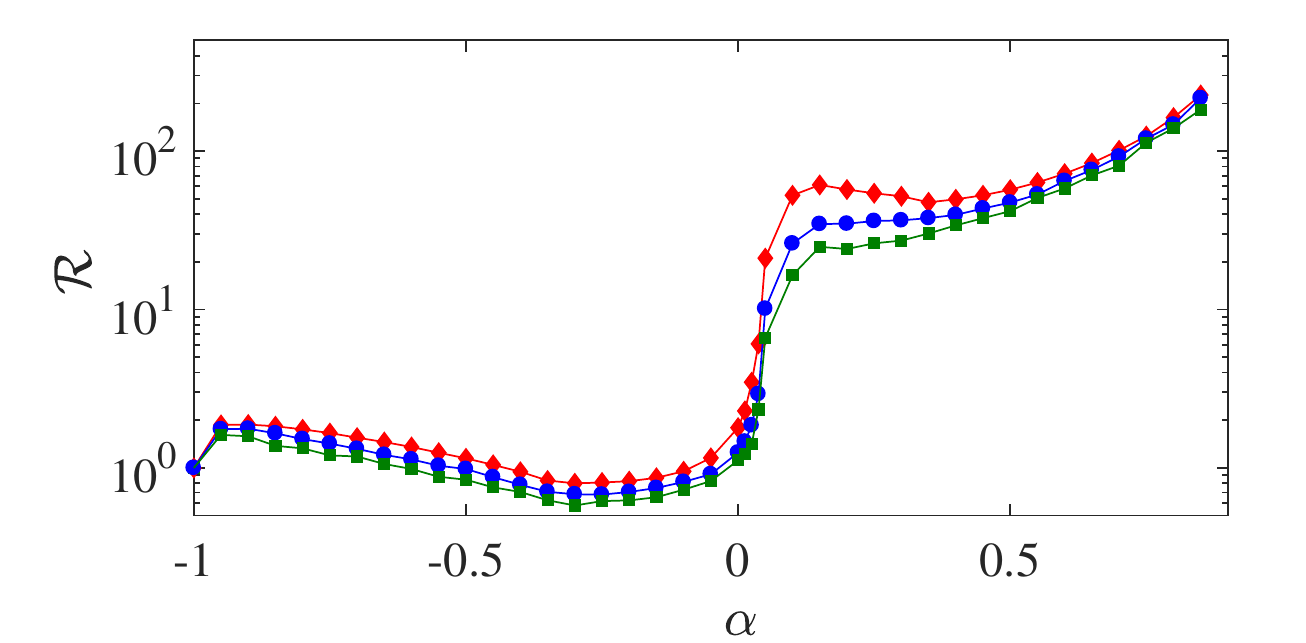}
\caption{ (a) Rectification $\Rec$ as a function of $\alpha$ for $\phi=\pi$ (red $\diamond$), and average rectification sample over $100$ samples of the phase picked from $\phi\in \pi+\pi/20[-1,1]$ (blue $\circ$) and $\phi\in \pi+\pi/10[-1,1]$ (green $\square$). Other parameters are $L=1000$, $\lambda=0.9$, $T_{h}=T_c+1000$ and $T_c=0.1$. }
\label{fig:Fig5} 
\end{figure}

It is important to show that the effect of the mobility edge on the rectification is robust to changes of the quasi-periodic potential phase parameter $\phi$. We thus study the rectification $\Rec$ versus $\alpha$ for $\phi=\pi$ (red $\diamond$) in Fig.~\ref{fig:Fig5}, and the average rectification for $\phi$ chosen from a uniform distribution between $\pi\pm\pi/20$ (blue $\circ$) and $\pi\pm\pi/10$ (green $\square$). The average rectification is computed taking $100$ samples of $\phi$. We observe that, while noise in the value of the phase $\phi$ lowers the rectification and makes the jump close to $\alpha=0$ less steep, the average rectification is robust even to variations of $10\%$ of the mean value.

\begin{figure}
\includegraphics[width=\columnwidth]{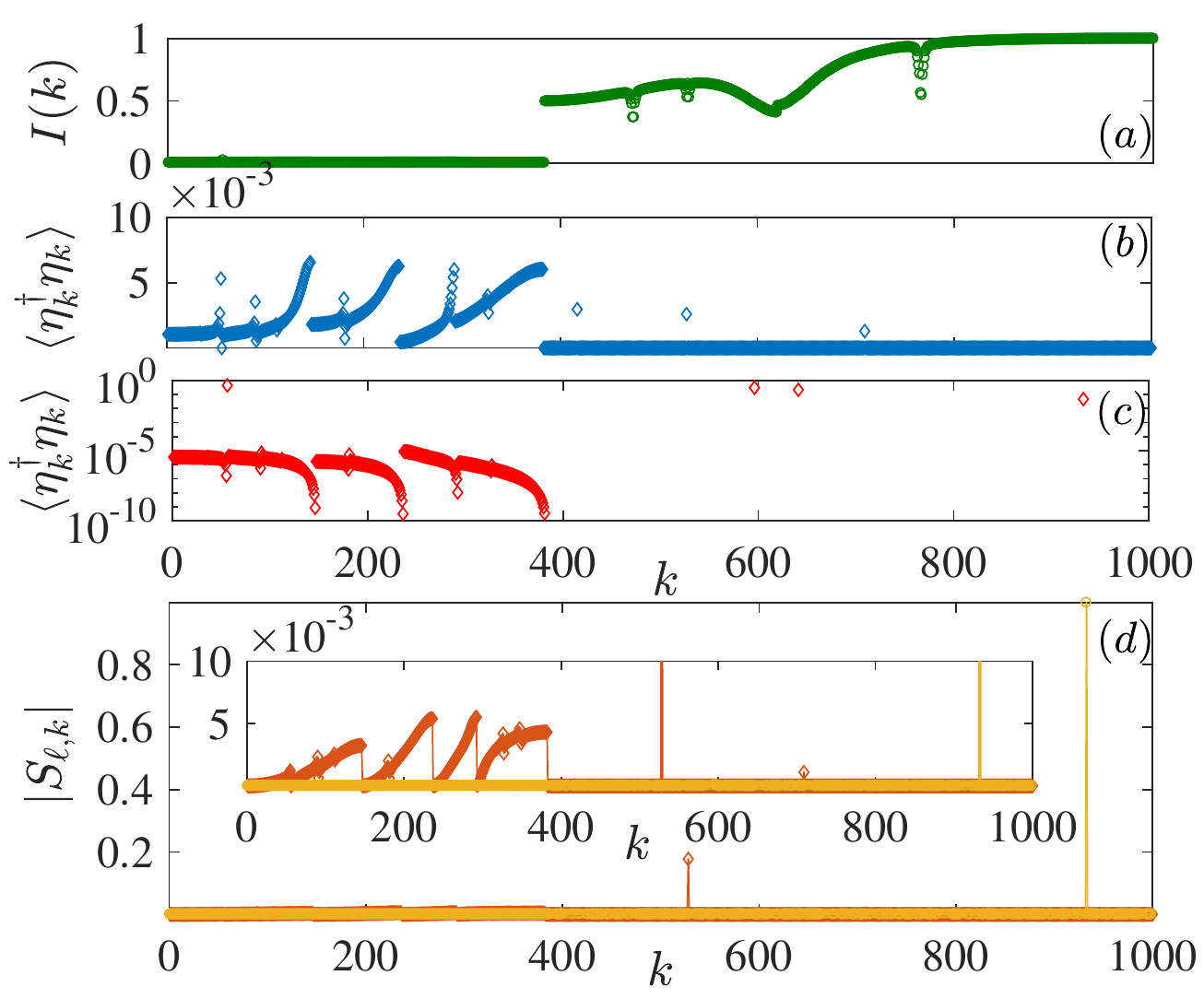}
\caption{(a) Inverse participation ratio $I(k)$ of the $k$th eigenmode of the Hamiltonian of Eq.~(1) of the main paper. (b)-(c) Occupation probability of the eigenmodes $k$, $\langle \eta_k^{\dagger}\eta_k \rangle$, versus the mode number $k$ ordered for increasing energy and for the (b) forward (in a lin-lin plot) and (c) reverse bias (in a log-lin plot). 
(d) Strength of the couplings  $|S_{\ell,k}|$ of eigenmodes $k$ to the left bath, i.e $\ell=1$ (red $\diamond$), and right bath, $\ell=L$ (yellow $\circ$). The inset magnifies a portion of the y-axis to better show the coupling to delocalized modes. Common parameters are $\phi=\pi$, $T_h=1000.1$, $T_c=0.1$, $L=1000$, $\lambda=0.9$ and $\alpha=0.9$.}
\label{fig:FigS2}
\end{figure}   
 
\section{Localization properties of the steady state}   

The link between rectification and localization is further unravelled by examining more closely a representative case with $\lambda = 0.9$ and $\alpha = 0.9$. In Fig.~\ref{fig:FigS2}(a) we report the inverse participation ratio $I(k)$ for the energy eigenmodes, highlighting the mobility edge above which $I(k)$ becomes sizeable indicating localized eigenstates. The NESS eigenmode occupation $\langle \eta_k^{\dagger}\eta_k \rangle$ for forward and reverse bias is shown in Fig.~\ref{fig:FigS2}(b)-(c). This reveals that in reverse bias most of the occupied eigenstates are the higher energy localized ones, while in forward bias there is significant occupation of lower energy delocalized eigenstates. 

A deeper insight into the generation of such strong rectification via the mobility edge can be obtained by studying the strength of the coupling of each bath to the various $k$ modes, i.e. $S_{\ell,k}$. In Fig.~\ref{fig:FigS2}(d) we show $|S_{L,k}|$ versus $k$ (yellow $\circ$) and $|S_{1,k}|$ (red $\diamond$). The inset is used to zoom in on the vertical axis so as to show the magnitude of $|S_{1,k}|$ for the delocalized modes. In Fig.~\ref{fig:FigS2}(d) we observe that for $\phi=\pi$ the bath at site $L$ is almost completely coupled only to one mode, which is localized at that edge. In fact there is a sharp peak for a high$-k$ mode which almost reaches unity. This also implies that the delocalized modes are very weakly coupled to this bath. For the bath at the first site, instead, $|S_{1,k}|$ is much more strongly coupled to delocalized modes. So the spatial position of the localized modes affects the strength of the coupling between a bath and the delocalized current-carrying modes. When connecting this with Eq.~(4) of the main paper, we observe that the interplay between the bosonic system and spin bath statistics, and the disparity in coupling strengths to the delocalized modes of the baths which together result in a significantly different current in the two biases.  

\begin{figure}
\includegraphics[width=\columnwidth]{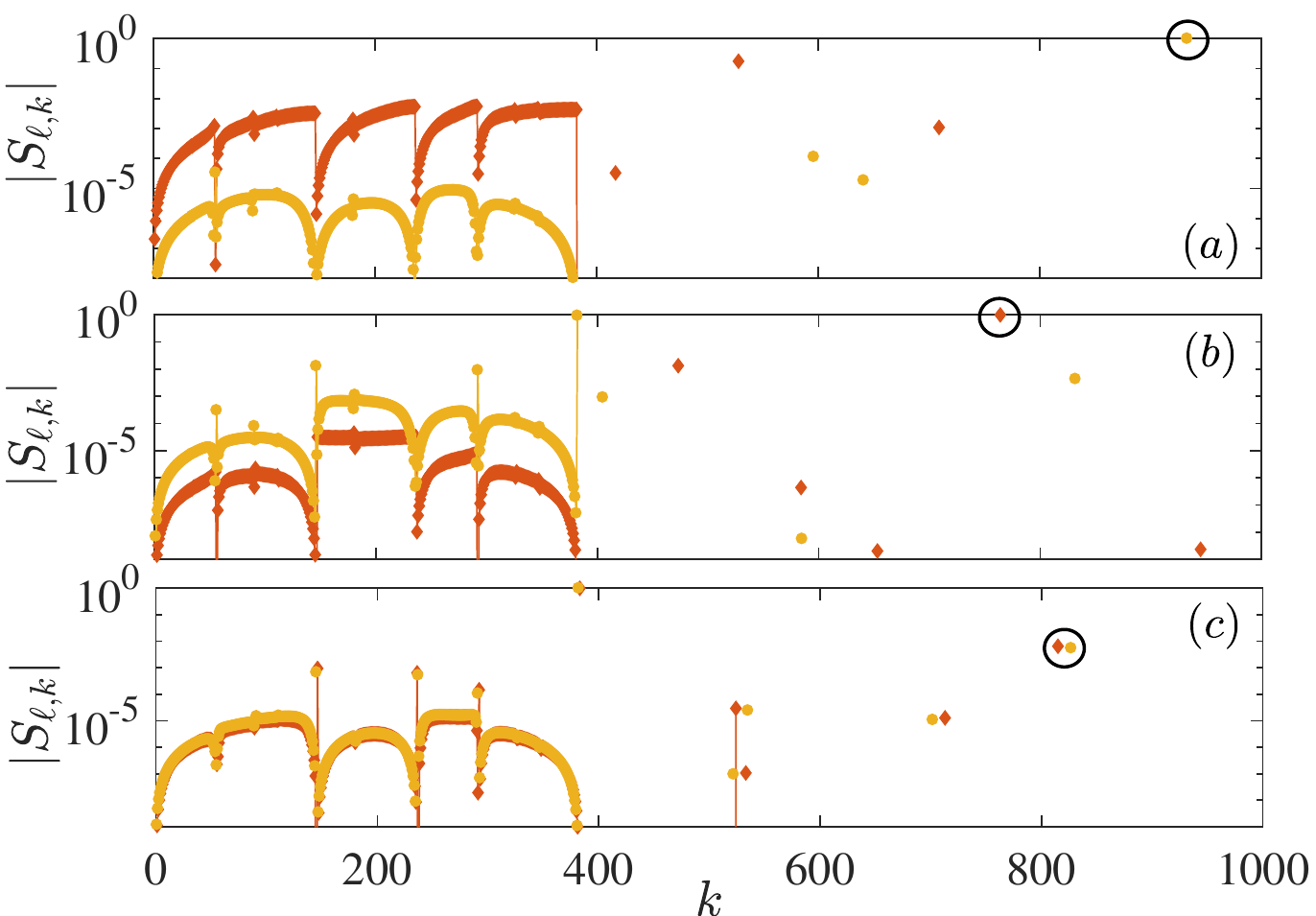}
\caption{(a-c) Strength of the couplings  $|S_{\ell,k}|$ of eigenmodes $k$ to the left bath, i.e $\ell=1$ (red $\diamond$), and right bath, $\ell=L$ (yellow $\circ$),  for (a) $\phi=\pi$, (b) $0$ (b) and (c) $1.07$. The forward current, when compared to the reverse current, is larger in (a), lower in (b) and comparable in (c). Common parameters are $\lambda=0.9$ and $\alpha=0.9$. The black circles highlight the more strongly coupled localized modes.}
\label{fig:FigS3}
\end{figure}

By tuning the phase $\phi$ it is possible to move the localized modes and hence modify which bath is coupled to a localized mode. In Fig.~\ref{fig:FigS3} we show the coupling of the modes to the two baths $|S_{\ell,k}|$ for (a) $\phi=\pi$, (b) $\phi=0$ and (c) $\phi=1.07$. The three panels show the couplings to the bath at site $\ell=1$ (red $\diamond$) and to the bath at site $\ell=L$ (yellow $\circ$). For Fig.~\ref{fig:FigS3}(a) the current is stronger in forward bias, and in fact the coupling to the delocalized modes is stronger for the bath at site $\ell=1$, while the bath at site $\ell=L$ there is a strong coupling to a single localized mode (highlighted by the large black circle) giving $\Rec \gg 1$. In Fig.~\ref{fig:FigS3}(b) the situation for the couplings is inverted, so now there is a strong coupling to a single localized mode for the bath at site $\ell=1$ (see within the large black circle) and the bath at site $\ell=L$ is more strongly coupled to delocalized modes giving $\Rec \ll 1$. In Fig.~\ref{fig:FigS3}(c) the coupling of localized and delocalized modes is similar for both baths, and it results that rectification is $\Rec\approx 1$.             
 
\section{Role of temperature in modes occupation}
\begin{figure}
\includegraphics[width=\columnwidth]{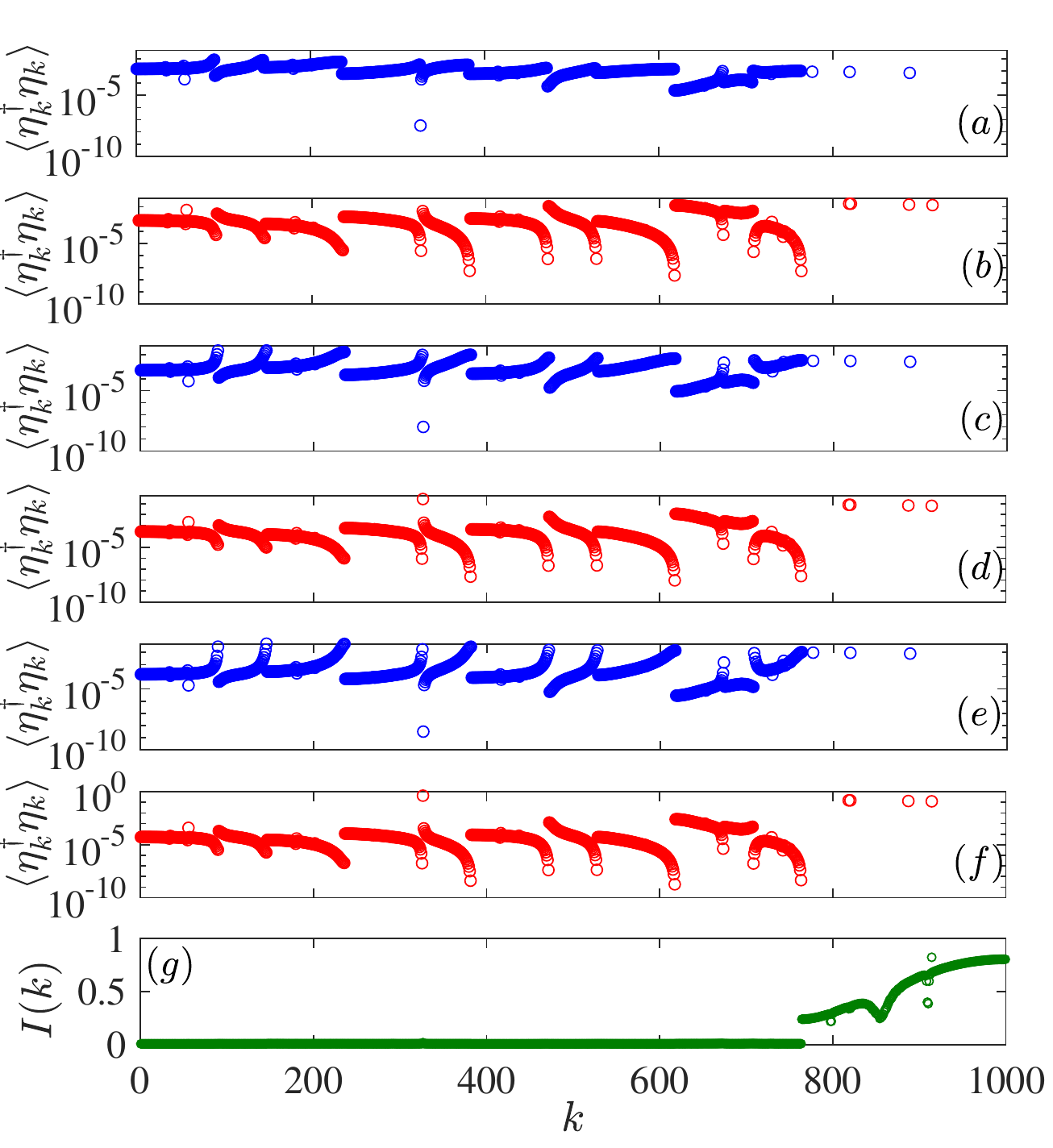}
\caption{ (a)-(f) Occupation probability of the eigenmodes $k$, $\langle \eta_k^{\dagger}\eta_k^{}\rangle$, versus the mode number $k$ ordered for increasing energy. We consider $T_h=10.1$ for the (a) forward and (b) reverse bias. We consider $T_h=100.1$ for the (c) forward and (d) reverse bias. We consider $T_h=1000.1$ for the (e) forward and (f) reverse bias. (g) Inverse participation ratio $I(k)$ of the $k$th eigenmode of the Hamiltonian. Common parameters are $L=1000$, $\phi=\pi$, $T_c=0.1$, $\lambda=0.1$ and $\alpha=0.9$.}
\label{fig:Fig1S}
\end{figure}

In Fig.~\ref{fig:FigS2} we have studied the occupation of the different $k$ eigenmodes of the Hamiltonian in the steady state for a temperature difference $\Delta T=1000$. Here we show how the occupation of the different modes is affected by the temperature difference. To show the generality of the effect we also consider different Hamiltonian parameters, namely $\alpha=0.9$ and $\lambda=0.1$. We study the occupation of all the modes $\langle \eta_k^\dagger\eta^{}_k \rangle$ in Fig.~\ref{fig:Fig1S} in forward (a),(c),(e) and reverse (b),(d),(f) bias. The localization of the modes is signalled by the modes' inverse participation ratio $I(k)$. For the parameters considered there is a clear transition between delocalized and localized modes around the mode $k=765$. An increase in the hot temperature $T_h$ is reflected in a larger occupation of the localized modes (which are at higher energy) and a lower occupation of the delocalized ones. This is particularly evident in reverse bias.

\section{Master equation and role of statistics of bath in rectification}
We consider the total Hamiltonian of system plus bath at sites $\ell = \{1,L\}$ as
\begin{align}
    H_T=H + \sum_{\ell=1,L} \left( H_{sS,\ell} +H_{S,\ell} \right),  \label{eq:HamT} 
\end{align} 
where $H$ is defined in the main paper, while the two spin baths are identical, except for the temperature $T_\ell$, and have Hamiltonian $H_{S,\ell} = \sum_{\nu} \varepsilon_{\nu}\sigma^z_{\nu, \ell}/2$, where $\sigma^z_{\nu,\ell}$ is the Pauli-$z$ operator for the $\nu$th spin in the bath at site $\ell$ and $\varepsilon_{\nu}$ is its associated energy gap. The Hamiltonian that couples the system to a spin bath is taken as $H_{sS,\ell}=\sum_{\nu, \ell}g_{\nu}(a^{}_\ell+a_\ell^\dagger)(\sigma^+_{\nu,\ell}+\sigma^-_{\nu,\ell})$, where $g_{\nu}$ is the coupling strength of $\nu$th spin in the bath to its respective boundary site $\ell$, assumed to be identical for both, and $\sigma^\pm_{\nu,\ell}$ are its corresponding raising and lowering operators.

In terms of the eigenoperators of the system Hamiltonian, we can write $H_{sS,\ell}=\sum_{\alpha \omega}A_{\ell,\alpha}(\omega) \otimes B_{\ell,\alpha}$, where $A_{\ell,\alpha}(\omega)$ and $B_{\ell,\alpha}$ respectively act on the system and on the bath. The operator $A_{\ell,\alpha}(\omega)$ is chosen to satisfy
\begin{eqnarray}
[H, A_{\ell,\alpha}(\omega)] &=& -\omega A_{\ell,\alpha}(\omega) 
\end{eqnarray} 
Note that, after rotating wave approximation, $\alpha$ takes two values with $A_{\ell,1}(\omega) = \sum_{k}S_{\ell,k}\eta_{k}\delta_{\omega,+\epsilon_{k}}$, $A_{\ell,2}(\omega) =\sum_{k} S^{*}_{\ell,k}\eta^{\dagger}_{k}\delta_{\omega,-\epsilon_{k}}$, while $B_{\ell,1} =\sum_{\omega}g_{\ell,\omega}\sigma^+_{\omega,\ell}$ and $B_{\ell,2} =\sum_{\omega}g_{\ell,\omega}\sigma^-_{\omega,\ell}$.
For the above thermal bath coupling, the steady state can be obtained from the Lindblad dissipator \cite{BreuerPetruccione2007}
\begin{eqnarray}\label{dissi}
   D_{\ell}(\rho (t))&=&\sum_{\omega, \alpha}\Gamma_{\ell, \alpha}(\omega)[A_{\ell, \alpha}(\omega)\rho (t) A_{\ell, \alpha}^{\dagger}(\omega)\nonumber \\
         &&\quad\quad-\frac{1}{2}\{A_{\ell, \alpha}^{\dagger}(\omega)A_{\ell, \alpha}(\omega),\rho (t)\}],
\end{eqnarray}
with
\begin{equation}\label{corrf}
  \Gamma_{\ell,\alpha}(\omega)=\int_{0}^{\infty} dt~ e^{i\omega t}\mathrm{Tr}_{B}[B^{\dagger}_{\ell, \alpha}(\tau)B_{\ell,\alpha}(0)\rho_{S,\ell}]. 
\end{equation} 
Here $\rho_{S,\ell}=\exp(-\beta_{\ell}H_{S,\ell})/Z$ is the thermal state of the bath coupled to site $\ell$ at inverse temperature $\beta_{\ell}=1/T_{\ell}$ and $Z={\rm tr}\left[\exp(-\beta_{\ell}H_{S,\ell})\right]$.  
Thus,
\begin{eqnarray}\label{corrf1}
  \Gamma_{\ell,1}(\omega) &=&J_\ell(\omega)(1-n_{S}(\beta_\ell\omega)).
\end{eqnarray}
where $n_S(\beta_\ell\omega)=\left(e^{\beta_\ell\omega}+1\right)^{-1}$ is the spin occupation factor for the bath coupled to site $\ell$, and $J(\omega)=\sum_{\nu}|g _{\nu}|^{2} \pi \delta(\omega-\varepsilon_{\nu})$ is the spectral density identical for both baths. Similarly
\begin{eqnarray}\label{corrf1}
  \Gamma_{\ell,2}(\omega) &=&J(-\omega)n_{S}(-\beta_\ell\omega).
  \end{eqnarray}
For any operator $O$ the dissipator evolution is
  \begin{eqnarray}
     D_{\ell}(O)&=&\sum_{\epsilon_{k} }|S_{\ell,k}|^{2} J(\epsilon_{k})\Big[n_{S}(\beta_\ell\epsilon_{k})\big[\eta_{k}O\eta_{k}^{\dagger}-\frac{1}{2}\{\eta_{k}\eta^{\dagger}_{k},O\}\big]\nonumber \\
     &&+ (1-n_{S}(\beta_\ell\epsilon_{k}))\big[\eta_{k}^{\dagger}O\eta_{k}-\frac{1}{2}\{\eta_{k}^{\dagger}\eta_{k},O\}\big]\Big ].
  \end{eqnarray}
  In the steady state limit, $\sum_{\ell}D_{\ell}(O)=0$, as expected.
   Thus, the steady state single-particle density matrix $\langle\eta_{k}^{\dagger}\eta_{q}\rangle $ is
    \begin{align}
     \langle \eta_{k}^{\dagger}\eta^{}_{q} \rangle &= \delta_{k,q}   \frac{\sum_{\ell}|S_{\ell,k}|^{2}n_S(\beta_\ell\epsilon_{k})}{ \sum_{\ell}|S_{\ell,k}|^{2}\left[1-2 n_S(\beta_\ell\epsilon_{k})\right]}.
\end{align}
     The thermal current is given by the energy exchange with each bath which in steady state needs to be opposite. Hence we get   
     \begin{eqnarray}
   \mathcal{J} &=& \mathrm{Tr}\{\mathrm{H} D_{1}(\rho)\} = - \mathrm{Tr}\{\mathrm{H} D_{L}(\rho)\}, \nonumber \\
   &=& \sum_{ \epsilon_{k} > 0}\epsilon_{k}|S_{1,k}|^{2}|S_{L,k}|^{2}J(\epsilon_{k})\nonumber \\
   &&\quad\quad\times\frac{n_S(\beta_1\epsilon_{k})-n_S(\beta_L\epsilon_{k})}{ \sum_{\ell}|S_{\ell,k}|^{2}(1-2n_S(\beta_\ell\epsilon_{k}))}.
   \end{eqnarray} 
   Suppose we use bosonic bath instead of spins with a different system-bath coupling term $H_{sB,\ell} = \sum_{\nu}g_{\nu}(a_\ell+a_\ell^\dagger)(b_{\nu,\ell}+b^{\dagger}_{\nu,\ell})$, in terms of bath bosonic creation $b^{\dagger}_{\nu,\ell}$ and annihilation $b_{\nu,\ell}$ operators for the $\nu$th mode of the bath at site $\ell$. In this case the bath Hamiltonian is given by $H_{B,\ell}=\sum_{\nu} \varepsilon_\nu \; b^\dagger_{\nu,\ell}b^{}_{\nu, \ell}$.   
   Following the above steps, we get  $\Gamma_{\ell,1}(\omega) = J(\omega)[1+n_B(\beta_\ell\omega)]$ and $\Gamma_{\ell,2}(\omega) = J(-\omega)n_B(-\beta_\ell\omega)$, where $n_B(\beta_\ell\omega)=\left(e^{\beta_\ell\omega}-1\right)^{-1}$ is the Bose-Einstein distribution. Hence, the occupation of each $k$ eigenmode when the system is coupled to the two bosonic baths is 
   \begin{align}
     \langle \eta_{k}^{\dagger}\eta^{}_{k} \rangle &=   \frac{\sum_{\ell}|S_{\ell,k}|^{2}n_B(\beta_\ell\epsilon_{k})}{ \sum_{\ell}|S_{\ell,k}|^{2}},
\end{align}
     and the steady state thermal current is
     \begin{eqnarray}
   \mathcal{J} &=& \mathrm{Tr}\{\mathrm{H} D_{1}(\rho)\} = - \mathrm{Tr}\{\mathrm{H} D_{L}(\rho)\}, \nonumber \\
   &=& \sum_{k}\frac{\epsilon_{k}|S_{1,k}|^{2}|S_{L,k}|^{2}J(\epsilon_{k})}{ \sum_{\ell}|S_{\ell,k}|^{2}}\nonumber \\
   &&\quad\quad \times[n_B(\beta_1\epsilon_{k})-n_B(\beta_L\epsilon_{k})].
   \end{eqnarray}
Thus, it follows that a bosonic system connected to bosonic baths gives no rectification, whereas a rectifying effect can be obtained using spin baths.

\end{appendix}


\begin{thebibliography}{100}
\bibitem{simplemodel} For a non-zero current both baths must couple to at least one common eigenstate $K$, which to have an overlap with both boundaries is necessarily delocalized. For the scenario described the bath at $\ell=L$ has a coupling dominated by a localized eigenstate while the one at $\ell=1$ is coupled to delocalized eigenstates. This is modelled by taking $S_{L,K}=q$ and $S_{1,K}=Q$ with $Q \gg q$. For $T_h = \infty$ and $T_c = 0$, for this eigenstate Eq.(\ref{curr}) gives $\cur_f = \frac{1}{2}\epsilon_K J(\epsilon_K)Q^2$ in forward bias and $\cur_r = -\frac{1}{2}\epsilon_K J(\epsilon_K)q^2$ in reverse bias. Thus $\Rec = (Q/q)^2 \gg 1$.
\end{thebibliography}
\end{document}